\documentclass[12pt]{amsart}
\usepackage[margin = 0.9in]{geometry}
\usepackage{graphicx}
\usepackage{subfig}
\usepackage{enumitem}
\usepackage{amsaddr}
\newcommand{\invchi}{\mathrm{Inv}\chi^2}
\begin{document}
\title[Partisan Lean of States]{Partisan Lean of States:\\ Electoral College and Popular Vote}

\author{Andrey Sarantsev}

\address{Department of Mathematics \& Statistics, University of Nevada, Reno}

\email{asarantsev@unr.edu}

\begin{abstract}
We compare federal election results for each state versus the USA in every second year from 1992 to 2018, to model partisan lean of each state and its dependence on the nationwide popular vote. For each state, we model both its current partisan lean and its rate of change, as well as sensitivity of state results with respect to the nationwide popular vote, using Bayesian linear regression. We apply this to simulate the Electoral College outcome in 2020, given even (equal) nationwide popular vote, as well as 2016, 2008, and 2004 nationwide popular vote. We backtest 2012 and 2016 elections given actual popular vote. Taking equal popular vote for two major parties, we prove that the Electoral College is biased towards Republicans.
\end{abstract}



\maketitle

\thispagestyle{empty}

\section{Introduction}

Numerous models aim to forecast nationwide popular vote (PV) in the USA presidential elections. They derive it from economic data, foreign policy, ethnic composition and other factors: the book \cite{CampbellBook} and articles \cite{4people, Campbell, Erikson, Graefe, Graefe18}. In the USA, uniquely among developed nations, the winner is decided by the Electoral College (EC), rather than PV; see \cite{ECReview}. 

The EC consists of 538 people, with each of the 50 states delegating the number equal to its House and Senate members combined, and Washington, District of Columbia (DC) delegating 3 members. (DC is not a state since it does not have representatives in the House and the Senate.) For each state and the DC, its members of the EC must vote for the winner of the popular vote in the state. (Exceptions are Maine and Nebraska; see Section 2.) The current two-party system of Democrats and Republicans dates back to the Civil War, and the third-party candidates have never won a presidential election; only on a few occasions, they won some EC votes. The candidate winning 270 or more Electoral College becomes the  president. If there is a tie, the decision is deferred to the House of Representatives. 

This peculiarity of election system in the USA moved to the forefront since 2000, which was the first time since 1888 when the winner of the Electoral College lost the popular vote (that is, got lower \% of votes than the other major-party candidate). Such possibility was discussed in the book \cite{Prophecy}, published in 1991. In 2016, the same situation repeated. Both times, the winners of the EC were Republicans. This stirred a debate in the media whether the Electoral College system currently has a built-in bias against Democrats. It is impossible to do full justice to this literature; let us mention as an example \cite{BadEC, Gelman, GelmanNew, Sort, GoodEC, SmallState, Warf, Zingher}. 

Another feature of the USA electoral system is the division of states into `red', `blue', and `purple'. A `red' state has a very high probability of being won by Republicans, so that its EC members will vote for Republicans by a large margin. A `blue' state is the opposite: It is very likely to be won by Democrats, and the margin is likely to be large. `Purple' states, otherwise known as `swing' or `battleground' states, have a significant probability of being won by either party. This classification is somewhat informal, and there are no generally accepted thresholds. In particular, political analysts often split states into more than three categories. An example is seven categories: Solid/Likely/Lean (Democrat or Republican), and Swing. 

In the election season, American and international media often feature maps of the USA with states colored in various shades of red, blue, and purple, as election prediction. For example, California is considered a solidly blue state: It was won by Democrats in every presidential election since 1992, and the margin is now overwhelming (over 20\%). 

Cook Political Report \texttt{www.cookpolitical.com} created a somewhat more formal version of this classification: {\it Cook Partisan Voting Index} based on the last two President election, by comparing statewide vs nationwide election results. The web site 538 \texttt{www.fivethirtyeight.com} uses a similar version of this index. However, we would like to make this research more formal and statistically rigorous. We use Presidential, House, and Senate elections.

We classify states into solid R, lean R, swing, lean D, and solid D. We define the partisan lean of each state and model its time evolution, as well as correlation with the nationwide popular vote percentages. We use not only presidential elections (which happen every 4 years in the USA), but House and Senate elections (which happen every 2 years). We collect publicly available data from the House Clerk, Federal Election Commission (FEC) and Wikipedia, starting from 1992. See \cite{Sort, Alignment} on political evolution of states. 

For each of these elections: House, Senate, Presidential, take $D$ and $R$, the numbers of votes by major parties, and compute the quantity $z = \ln(D/R)$ for each state. Regress it upon the nationwide $\ln(D/R)$, and the year of election. We obtain both point estimates and Bayesian posterior, given a non-informative Jeffrey's prior. We simulate the 2020 EC given four different scenarios of PV: (a) Even PV, $D = R$; (b) 2016 PV, $D = 48.2\%, R = 46.1\%$; (c) 2008 PV, $D = 52.9\%, R = 45.7\%$; (d) 2004 PV, $D = 48.3\%, R = 50.7\%$.

We also simulate past elections in 2012 and 2016, given even PV. This allows us to see whether the EC is biased towards Republicans. That is, whether win probability for Democrats is significantly less than 50\%. This is true for 2012, 2016, and 2020. Finally, we investigate which swing states are more important for winning, given each scenario. That is, which of them can be most easily swung given additional resources. We stress that our research is not a prediction of 2020 elections; for this, we need to use polls, fundraising, endorsements, and economic indicators.  Among books on this topic, see \cite{Silver}.

\subsection{Organization} In Section 2, we describe the data collection and organization, including some special data points which we had to modify for consistency. In Section 3, we introduce our linear regression model. Section 4 is devoted to fitting Bayesian linear regression for each state; and Section 5 contains simulations of the Electoral College and related discussions. Section 6 concludes the article and proposes future research.  

\subsection{Conflict of interest statement} On behalf of all authors of this manuscript, the corresponding author states that there is no conflict of interest.

\subsection{Acknowledgements} I am thankful to my undergraduate students Jaucelyn Canfield and Franklin Fuchs for collecting the data and helping me code the program. I am also thankful to my undergraduate student Akram Reshad for pointing to me the data source: FEC, which made data collection much easier. I would also like to thank Professors Aleksey Kolpakov and Thomas Kozubowski for useful discussion and pointing relevant literature. 

\section{Election Data}

\subsection{Data description} We collect statewide \% of popular vote for each of the two major parties, for each of the 50 states for each presidential, House, and Senate election since 1992. 

A House election happens every two years. Each state is split into several congressional districts, for each of which a representative for the House is elected for two years. The state with the most districts is California (53). Several low-populated states (Alaska, Wyoming, and others) have only one congressional district: {\it district-at-large}. In other states, districts are numbered: California 1, \ldots, California 53; and similarly for other states. 

For each state, we sum votes in all congressional districts of this state, for each major party. Then we divide these two numbers by the overall popular vote (for all major and minor parties together) in this state. The quantity of such congressional districts and their shape is determined by the population of the state relative to the USA. Each 10 years, after Census, the number and the shape of these districts are recalculated. In particular, after the coming 2020 Census, Texas is projected to gain 3 congressional districts, due to large growth in population.

Every state has two senators, elected statewide. Each senator has a six-year term. All 100 senators in the Senate are split evenly into three so-called Classes: Class I, Class II, and Class III, which determines their election years. This means that each state has Senate elections every two out of three even years. For example, California had Senate elections in 2016 and 2018, but will not have them in 2020. 

The President of the USA is elected every four years: 1992, 1996, 2000, etc. As described in the Introduction, each state is assigned EC members, equal in number to the total sum of House and Senate members. For the 48 states other than Maine and Nebraska, these EC members vote for the popular vote winner in this state. 

There is an issue of {\it faithless electors}, which break this rule. However, so far there were only very few such electors, and this has not influenced the outcome. Thus we shall not attempt such modeling in this article.

Maine and Nebraska use a hybrid system: They assign two EC votes to the statewide winner, and other EC votes to winners of congressional districts. This can lead to splits, for example 2008, Nebraska 2 vs the state of Nebraska; 2016, Maine 2 vs the state of Maine. However, the boundaries of the districts change with each redistricting, and thus we cannot compare the same district in different years. For the purposes of this article, we simply assume that Maine and Nebraska assign all EC votes to the statewide winner. This will introduce an error to our analysis, but it is of order 1 electoral vote, which is not much. 

Finally, we have a benchmark: the nationwide popular vote. For a Presidential election, this is self-explanatory. For a House election, this is the sum of votes in all 435 congressional districts. A Senate election is not nationwide (as explained above, each state has Senate elections 2 out of 3 even years). Thus we use the nationwide House PV of the same year. 

We do not model DC: It voted strongly Democratic in recent elections.

\subsection{Data, code, and results on GitHub}
The \texttt{GitHub} repository of the author, registered as user \texttt{asarantsev}, named  \texttt{Electoral-College-Vs-Popular-Vote}  contains data in \texttt{data.xlsx}: the sheet \texttt{RawData} contains numbers or \% of votes $D$ and $R$ in each election; the sheet \texttt{Logs} contains $\ln(D/R)$ for individual elections; the sheet \texttt{Means} contains averages of these logarithms for each state-year; and the sheet \texttt{EC} contains EC votes in each state, as well as state population in 2000 and 2010 Census, and the Cook Partisan Voting Index (we compare it with our own version of partisan lean index). This GitHub repository also contains the source code \texttt{code.py} and the output in the files \texttt{regressionResults.csv}, \texttt{stateResults.csv}, \texttt{importance.csv}.

\subsection{Data sources} Data for 2000--2016 is taken from the Federal Election Commission:\\ 
\texttt{https://transition.fec.gov/pubrec/electionresults.shtml} 

\begin{itemize}
\item 2004, 2008, 2012, 2016: Tables 2, 6, 7; 
\item 2006, 2010, 2014: Tables 4, 5;
\item 2002: Tables 2, 3;
\item 2000:  Sheets 2, 4, 5. 
\end{itemize}
For 1992--1998, the data is taken from the House Clerk web page:

\begin{itemize}

\item \texttt{http://clerk.house.gov/member\_info/electionInfo/1998/Table.htm}

\item \texttt{http://clerk.house.gov/member\_info/electionInfo/1996/Table.htm}

\item \texttt{http://clerk.house.gov/member\_info/electionInfo/1994/94Recapi.htm}

\item \texttt{http://clerk.house.gov/member\_info/electionInfo/1992/92Recapi.htm}
\end{itemize}

\subsection{Special elections} For some elections, we have to modify data.

\begin{enumerate}
\item A top-two primary system has two top vote-getters in the primary advance to the general election, regardless of the party.
Such system is used in Washington House and Senate elections since 2008, and in California House and Senate elections since 2014. In 2016 and 2018 California Senate elections, this led to both Democrats as general election candidates in Califorina. For these Senate races, we use the primary election, summing votes for all candidates from the two major parties. 

\item For House races, same-party runoffs happened, too, but for less than half of districts in each state, for each election. Since the general election has a much higher turnout than the primary election, we consider the general election in this case more representative. In this research, we set the following rule: We ignore a House election if in at least half of districts there was no candidate from either Democrats or Republicans.  

\item Louisiana has a system similar to California and Washington: There, the November election takes the form of a {\it jungle primary:} All candidates run together, not separated by party. If no candidate gets $50\%$, a runoff election is in December. We sum votes for all candidates from the two major parties in both the jungle primary and the runoff, if it happens, and divide these by the total vote. 

\item We ignore Louisiana House elections for 1996, 1998, 2002, 2012, 2014, and 2016: In each of these years, in at least half of districts, runoff elections were one-party.  

\item We treat Bernie Sanders from Vermont and Angus King of Maine as Democrats. Bernie Sanders took part in House elections for Vermont at-large district in 1992--2004 and in Senate elections for Vermont in 2006, 2012, and 2018. Angus King took part in Senate elections for Maine in 2012 and 2018. For each of these elections, we sum the votes of Bernie Sanders or Angus King and a Democrat in the same race, if such Democrat existed; and assign this percentage to Democrats. 
\end{enumerate}

There are some other elections which we have to ignore. 

\begin{enumerate}
\item 2016 House election in Vermont at-large had Peter Welch, for both major parties: He is a Democrat, but won the Republican primary on write-in votes. 

\item 2006 Senate election in Connecticut featured incumbent Joe Lieberman running (and winning) as an independent, because he lost the Democratic primary. Same applies to Lisa Murkowski in 2010, Alaska Senate race: She lost the Republican primary. 

\item The following Senate elections did not have a Democratic: 2010, South Dakota; 2002, Virginia; 2002, Mississippi;  2002, Virginia; 2006, Indiana; 2000, Arizona; 2014, Alabama; 2004, Idaho; 2014, Kansas. In each case, there was no opposition, or other candidates were not ideologically similar to Democrats. 

\item The following Senate elections did not have a Republican: 2002, Massachusetts; 2008, Arkansas.  In each case, there was no opposition, or other candidates were not ideologically similar to Republicans.

\item The following House elections did not have a Republican candidate for at least half of the districts in the state: 2008,  Vermont at-large; 2008, Arkansas 1, 2, 4;  2006, Rhode Island 2; 1996 and 1998, West Virginia 1, 3; Massachusetts, 2000--2008, 2014, 2016. 

\item The following House elections did not have a Democratic candidate for at least half of the districts in the state: 2016,  Arkansas 1, 3, 4; 1998, Nevada 2; 2012, Kansas 1, 3; 2002, Nebraska 1, 3.
\end{enumerate}

\begin{figure}
\centering
\subfloat[California]{\includegraphics[width = 5cm]{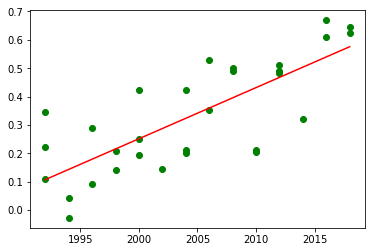}}
\subfloat[Nevada]{\includegraphics[width = 5cm]{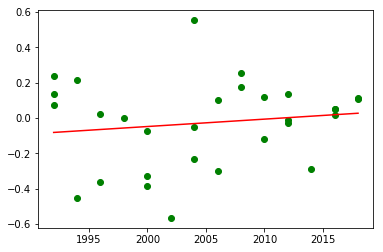}}
\subfloat[Texas]{\includegraphics[width = 5cm]{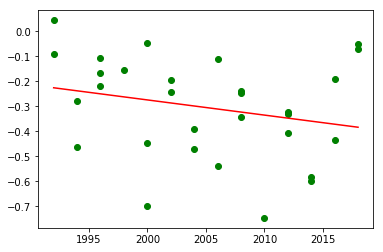}}
\caption{President, House, Senate elections: $\ln(D/R)$ for D and R votes. This regression differs from~\eqref{eq:basic}: We do not regress upon the nationwide PV.}
\label{fig:states}
\end{figure}

\section{Regression Model}

We denote election years 1992--2018 by $t = -14, \ldots, -2, -1$, to make the coming election year (2020, as of this writing) to be $t = 0$. We consider two approaches: Separating or averaging elections with same state and year.

\subsection{Separate elections} Let $S$ be the set of 50 states. Let $E$ be the set of all 35 House, Senate, and President elections. For each election $e \in E$, we denote its year by $t(e)$, and the party vote percentages nationwide by $d(e)$ and $r(e)$. Recall that we use House nationwide popular vote percentages for the corresponding Senate election. For each state $s$, we denote elections used in this state by $E_s$. For a state $s$ and an election $e \in E_s$, we denote by $d_{s}(e)$ and $r_s(e)$ the statewide number of votes for Democrats and Republicans. Then
$$
x(e) = \ln\frac{d(e)}{r(e)}\quad \mbox{and}\quad y_s(e) = \ln\frac{d_s(e)}{r_s(e)}.
$$
We consider the following linear regression:
\begin{equation}
\label{eq:basic}
y_s(e) = \alpha_s + \beta_sx(e) + \gamma_st(e) + \sigma_s\varepsilon_s(e),\quad \varepsilon_s(e) \sim \mathcal N(0, 1)\quad \mbox{i.i.d.}\quad e \in E_s.
\end{equation}
Here, $\alpha_s$ is the {\it current partisan lean}, $\beta_s$ is {\it elasticity} of the state: its responsiveness to changes in the national environment (measured by $x(e)$), and $\gamma_s$ is {\it partisan lean rate} of increase or decrease. We borrow terms from quantitative finance: A stock's  {\it alpha} is the excess return compared with the whole market, and its {\it beta} is its sensitivity to changes in the whole market. 

However, this model does not fit: In \texttt{output.csv} we computed $p$-values for Shapiro-Wilk normality test for regression residuals of each of the 50 states; and 21 of them have $p < 5\%$. Therefore the residuals are not normal (5\% of 50 is 2.5, which is far less than 21). This is a deficiency of this model.

\subsection{Election aggregation} Instead, let us average all logarithms of ratios for each elections in a given state in a given year. Denote the result by $y_{st}$ for state $s \in S$ and year $t = -14, -13, \ldots, -2, -1$ (corresponding to 1992, 1994, \ldots, 2018). For example, Washington in 1992 had 3 elections: Presidential, House, and Senate, with $y_s(e)$ equal to $0.306, 0.160, 0.305$. Their average is $y_{st} = 0.257$. Oregon in 2018 had only one election: House, with $y_s(e) = 0.413$. Thus the average is equal to the same number: $y_{st} = 0.413$. Similarly, for every fourth year starting from 1992, we average $x(e)$ for the House and Presidential election to get nationwide partisan lean $x_{t}$. For every fourth year starting from 1994, we simply take $x(e)$ for the House election as the nationwide partisan lean. We modify~\eqref{eq:basic} as follows:
\begin{align}
\label{eq:modified}
y_{st} = \alpha_s + \beta_sx_t + \gamma_st + \sigma_s\varepsilon_{st},\quad 
\varepsilon_{st} \sim \mathcal N(0, 1)\quad \mbox{i.i.d.}
\end{align}
These are 50 linear regressions. Total $695  = 14\cdot 50 - 5$ data points, because Louisiana 1994, Massachusetts 2002, Vermont 2002, West Virginia 1998 and 2018 did not have any elections. 

To test goodness-of-fit, we compute the Shapiro-Wilk normality test $p$-value for residuals; total 50 $p$-values. We use $\chi^2$ test to verify whether these $p$-values are uniformly distributed on $[0, 1]$, by splitting into 20 equal subintervals. We get $p = 0.63$. This confirms that the improved model~\eqref{eq:modified} fits the data.  We remark in passing that combining all 695 residuals in one array and applying Shapiro-Wilk test gives us $p < 0.01$. Thus these residuals are not i.i.d. normal.

\subsection{Election simulation} We simulate national elections. We do predictions for year 2020, $T = 0$, as well as back-simulations for years 2012 ($T = -4$) and 2016 ($T = -2$). We fix nationwide PV percentages $d$ and $r$. Democrats win state $s$ if 
\begin{equation}
\label{eq:reg-res}
\alpha_s + \beta_s\ln\frac{d}{r} + \gamma_sT + \sigma_s\varepsilon_s = y_s = \ln\frac{d_s}{r_s} > 0. 
\end{equation}
This has probability $p_s := \Phi\left(\sigma_s^{-1}(\alpha_s + \beta_sx + \gamma_sT)\right)$, where  $\Phi$ is the CDF of $\mathcal N(0, 1)$:
$$
\Phi(u) := \frac1{\sqrt{2\pi}}\int_{-\infty}^ue^{-z^2/2}\,\mathrm{d}z.
$$
Then we simulate each state, independently of others, and sum the corresponding EC votes of Democrats. Repeating this simulation many times, we get the distribution of EC votes. This, in turn, gives us the probability that Democrats win (getting more than 269 EC votes). 

\section{Parameter Estimates} 

\subsection{Point estimates} For each state $s \in S$, estimates of $\alpha_s,\, \beta_s,\, \gamma_s,\, \sigma_s$ from~\eqref{eq:basic}, number $N_s$ of elections, and current EC votes $\mathrm{EC}_s$, are in \texttt{output.csv}. The summary in 2020 is below. 

\begin{itemize}
\item The reddest state, measuring by $\alpha$, is Wyoming, $\alpha = -1.05$.
\item The bluest state is Hawaii, $\alpha = 0.967$.
\item The most neutral state (with $\alpha$ closest to $0$) is Wisconsin, $\alpha = 0.0029$.
\item The most rapidly blueing state is Vermont, $\gamma = 0.037$. 
\item The most rapidly reddening state is North Dakota, $\gamma = -0.136$. 
\item The state with least change rate is Kansas, $\gamma = -0.0007$.
\item The state which is most sensitive to the national environment is Nebraska, $\beta = 1.819$. 
\item The state which is the least sensitive is Rhode Island, $\beta = 0.013$
\item The only state with negative sensitivity is Louisiana, $\beta = -0.249$.  
\item The states with the highest and lowest $\sigma$ are North Dakota and California, respectively. 
\end{itemize}

\subsection{Size effect} The correlation coefficients between $\ln\sigma_s$ and the logarithm of the state population in years 1990, 2000, 2010 are equal to $-65\%, -67\%, -68\%$, respectively. This indicates that smaller states, on average, have higher standard error of regression than larger states. This is similar to the stock market, where smaller stocks have higher volatility. 

\subsection{Comparison with Cook Partisan Voting Index} The current (as of 2020) state partisan lean is $\alpha_s$ for the state $s$. We did linear regression of the list of these 50 $\alpha_s,\, s \in S$, vs Cook Partisan Voting Index (based on the last two presidential elections, 2012 and 2016). This dependence is very strong, with correlation $98\%$.  However, we believe that $\alpha$ is a better index, since it includes more data than from the two last presidential elections. 

\subsection{Bayesian regression} However, each state has only few observations, at most 14. Thus we cannot assume that these estimates of $\alpha$, $\beta$, and $\gamma$ for each state are very precise. Thus we use Bayesian linear regression. For background, see the textbook \cite{Hoff}; we also found useful \cite[Chapter 4]{Finance}. Assume a {\it non-informative prior} for each state $s \in S$: 
\begin{equation}
\label{eq:prior}
\pi_s(\alpha_s, \beta_s, \gamma_s, \sigma^2_s) \propto \sigma^{-2}_s.
\end{equation}
The symbol $\propto$ stands for ``proportional''. The term {\it non-informative} refers to the fact that we do not use any information (such as empirical mean) in this distribution~\eqref{eq:prior}. This prior is {\it improper:} That is, the integral of this density over all possible values is infinite. This seems to be a contradiction in terms: A probability distribution must integrate (or sum, if it is a discrete distribution) to $1$. But we can still use this prior from~\eqref{eq:prior} in Bayesian analysis. 

The next step is to compute the {\it  likelihood} $L(y_{st},\, t = -14, \ldots, -1,\,\mid \alpha_s, \beta_s, \gamma_s, \sigma_s^2)$. This is a product of Gaussian densities. We get posterior distribution from the Bayes' formula:
\begin{align}
\label{eq:Bayes}
\begin{split}
p&(\alpha_s, \beta_s, \gamma_s, \sigma^2_s\mid y_{st},\,t = -14, \ldots, -1) 
\\ & \propto L(y_{st},\, t = -14, \ldots, -1\mid \alpha_s, \beta_s, \gamma_s, \sigma_s^2)\cdot \pi_s(\alpha_s, \beta_s, \gamma_s, \sigma^2_s).
\end{split}
\end{align}
From the choice of a prior in~\eqref{eq:prior}, the posterior distribution is already known explicitly, see for example \cite[Chapter 9]{Hoff} or \cite[(4.8)--(4.10)]{Finance}. 

We introduce some notation: The inverse $\chi^2$ distribution $\invchi_n(c)$ with $n$ degrees of freedom, scale parameter $c$, and density (with $\Gamma(\cdot)$ the Gamma function):
$$
f(x) = \frac1{\Gamma(n/2)}\left(\frac n2\right)^{n/2}c^n\cdot x^{-n/2-1}\exp\left(-\frac{nc}{2x}\right);
$$
The multivariate $d$-dimensional normal distribution $\mathcal N_d(\mu, \Sigma)$ with mean $\mu$ and covariance matrix 
$\Sigma$, with density
$$
f(x) = \left(2\pi\det(\Sigma)\right)^{-d/2}\exp\Bigl(-\frac12(x - \mu)^T\Sigma^{-1}(x - \mu)\Bigr);
$$
The number $n_s$ of elections for state $s$; for almost all states, $n_s = 14$, but for a few of them, it is 12 or 13; and, finally, 
a $3\times 3$ matrix $M_s = \begin{bmatrix}\mathbf{1} & \mathbf{x} & \mathbf{t}\end{bmatrix}$, where $\mathbf{1}$ is a vector of $n_s$ unit numbers, $\mathbf{x}$ is the vector $y_{st}$ for the state $s$, and $\mathbf{t} = [-14, \ldots, -1]$ is the vector of years (with missing years if necessary).

\begin{figure}
\centering
\subfloat[$\alpha_s$]{\includegraphics[width = 5cm]{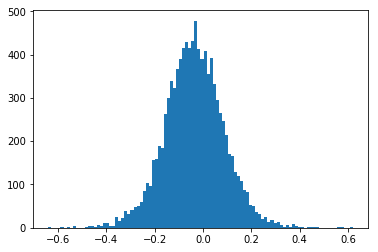}}
\subfloat[$\beta_s$]{\includegraphics[width = 5cm]{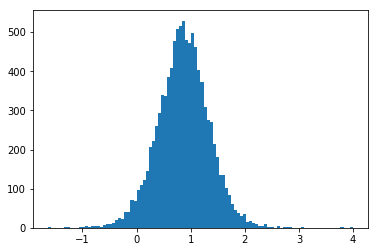}}
\subfloat[$\gamma_s$]{\includegraphics[width = 5cm]{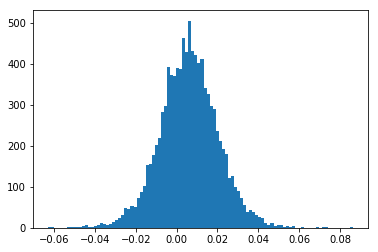}}
\caption{Histograms of $\alpha_s, \beta_s, \gamma_s$ for Nevada} 
\label{fig:state-hist}
\end{figure}

Then the posterior distribution from~\eqref{eq:Bayes}, as known from \cite{Hoff, Finance}, is 
\begin{align}
\label{eq:posterior}
\begin{split}
p(\alpha_s, \beta_s, \gamma_s\mid \sigma^2_s) &= \mathcal N_3((\hat{\alpha}_s, \hat{\beta}_s, \hat{\gamma}_s), \hat{\sigma}_s^2(M_s^TM_s)^{-1}),\\
p(\sigma^2_s) &= \invchi_{n_s-3}(\hat{\sigma}_s^2).
\end{split}
\end{align}
One can show that the posterior (unconditional) marginal distribution of $(\alpha_s, \beta_s, \gamma_s)$ is multivariate Student, with heavy tails. Simulated parameters $\alpha, \beta, \gamma$ for Nevada are given in  Figure~\ref{fig:state-hist}. 

\section{Electoral College Simulations}

We simulate each $\sigma_s$, then $\alpha_s, \beta_s, \gamma_s$, for each of the 50 states $s \in S$. Then we simulate the nationwide result in the following 6 scenarios: 
\begin{enumerate}[label=(\Alph*)]
\item 2020 election with even PV: $D = R = 50\%$;
\item 2020 election with 2016 PV: $D = 48.2\%,\, R = 46.1\%$;
\item 2020 election with 2008 PV: $D = 52.9\%,\, R = 45.7\%$;
\item 2020 election with 2004 PV: $D = 48.3\%,\, D = 50.7\%$.
\end{enumerate}

\begin{figure}
\centering
\subfloat[2020, PV even, $P = 37\%$]{\includegraphics[width = 7cm]
{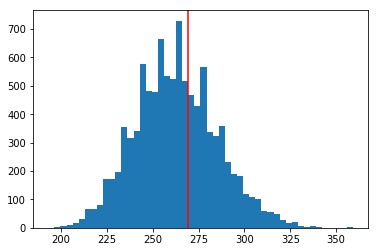}}
\subfloat[2020, PV 2016, $P = 66\%$]{\includegraphics[width = 7cm]
{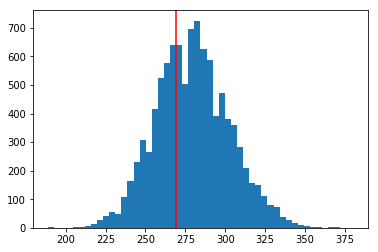}}
\\
\subfloat[2020, PV 2008, $P = 99\%$]{\includegraphics[width = 7cm]
{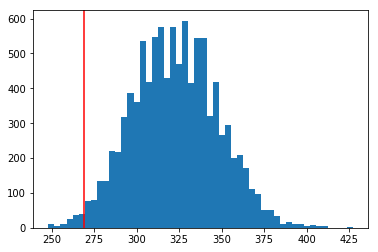}}
\subfloat[2020, PV 2004, $P = 13\%$]{\includegraphics[width = 7cm]
{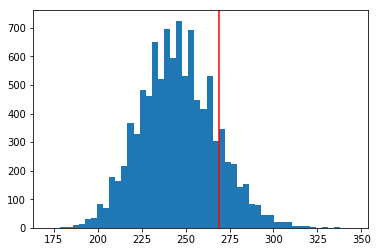}}
\caption{Democrats' EC votes and win probability $P$, given nationwide \% PV} 
\label{fig:EC}
\end{figure}

\begin{figure}
\centering
\subfloat[2020 with even PV]{\includegraphics[width = 3.2in, trim = 3in 2in 2in 2in]{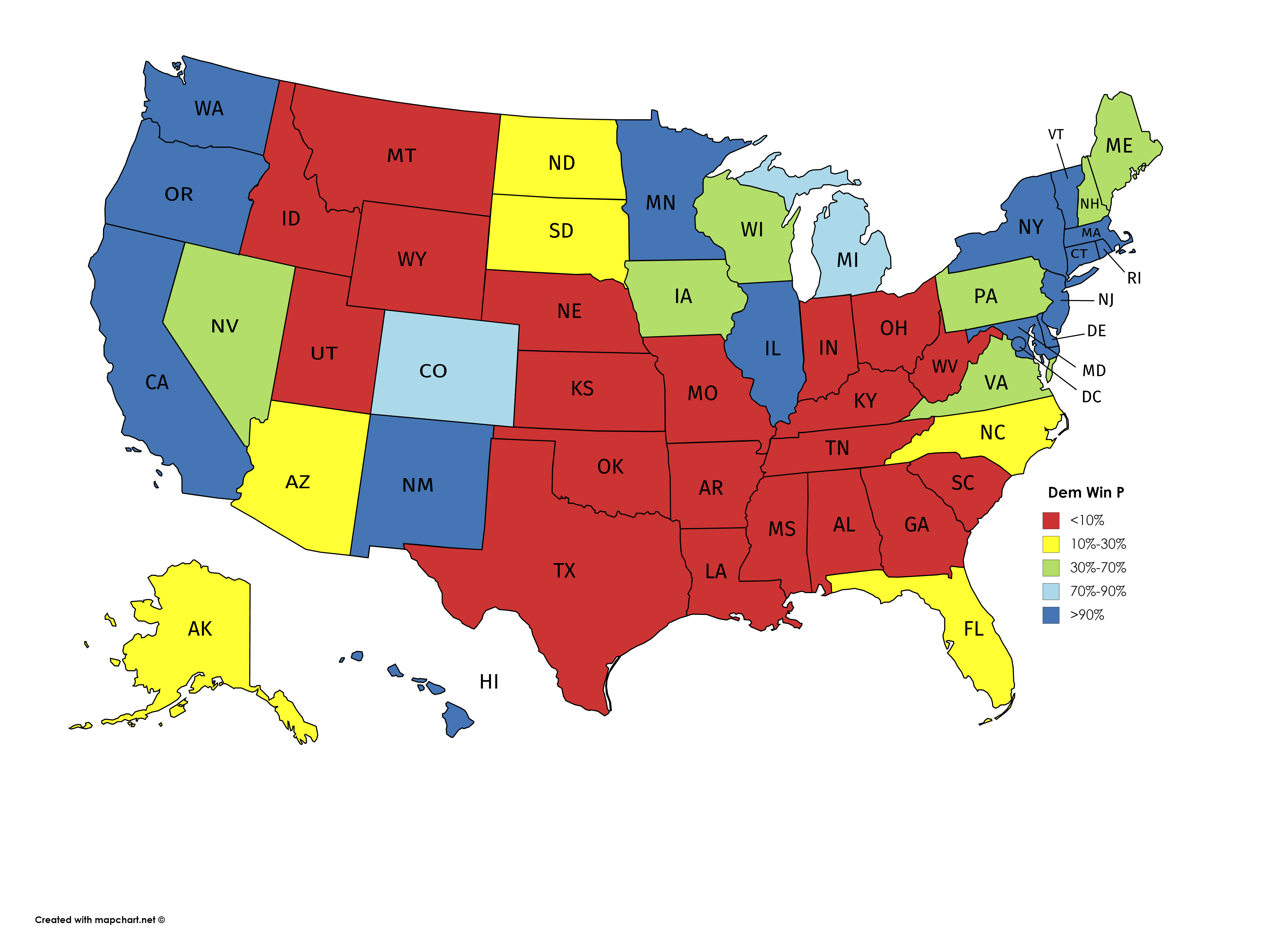}}
\subfloat[2020 with 2016 PV]{\includegraphics[width = 3.2in, trim = 3in 2in 2in 2in]{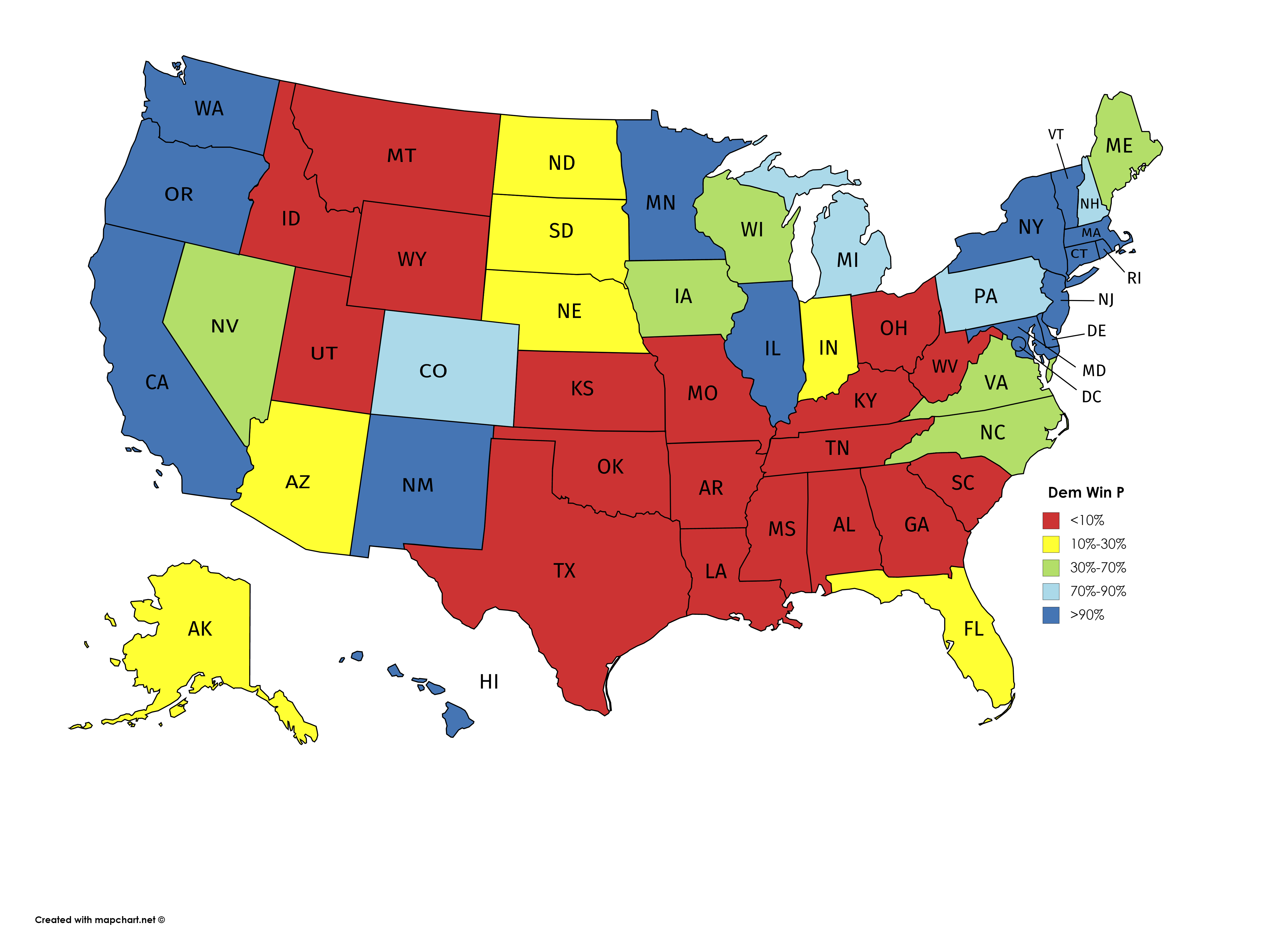}}
\\
\subfloat[2020 with 2008 PV]{\includegraphics[width = 3.2in, trim = 3in 2in 2in 2in]{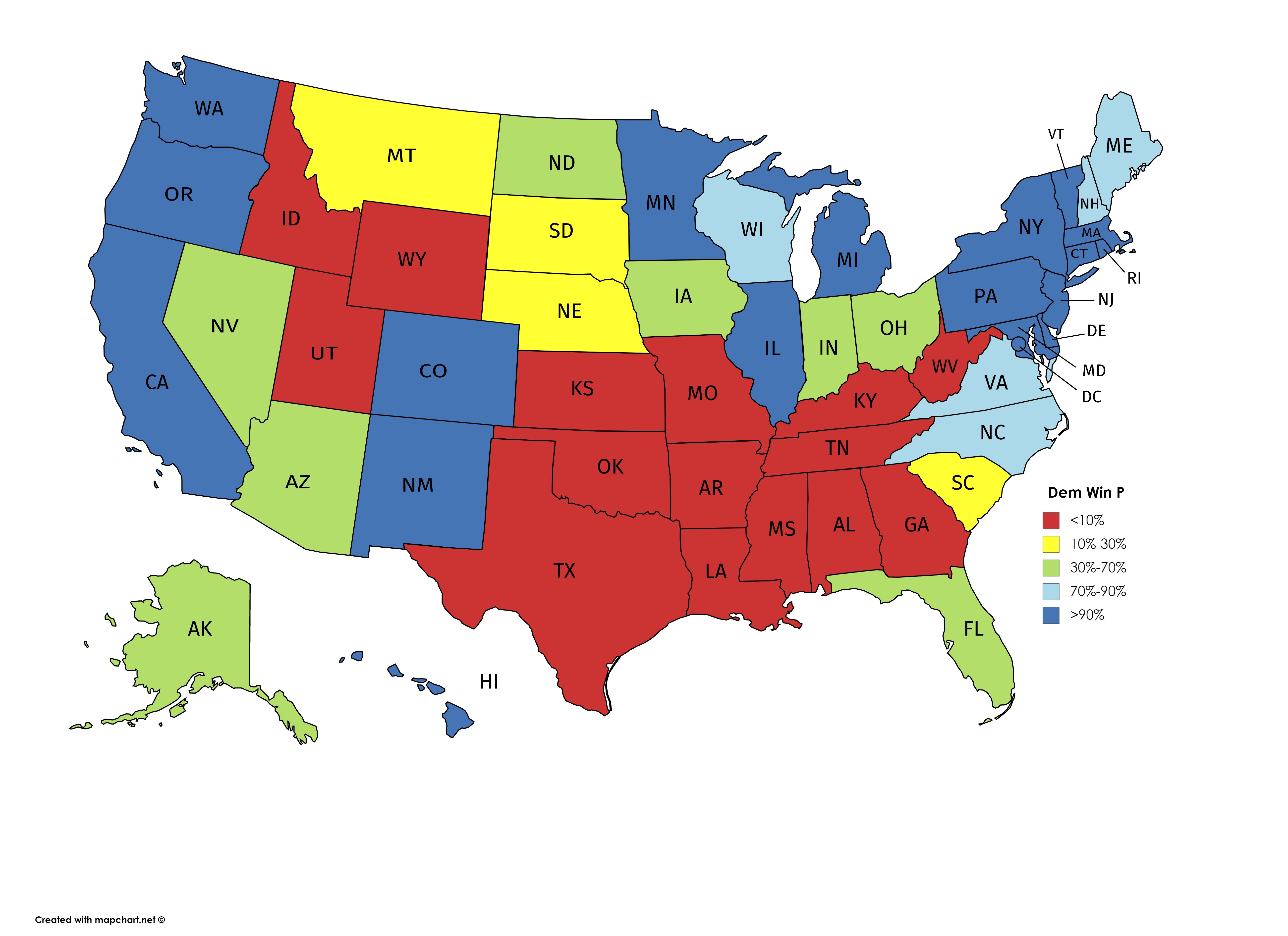}}
\subfloat[2020 with 2004 PV]{\includegraphics[width = 3.2in, trim = 3in 2in 2in 2in]{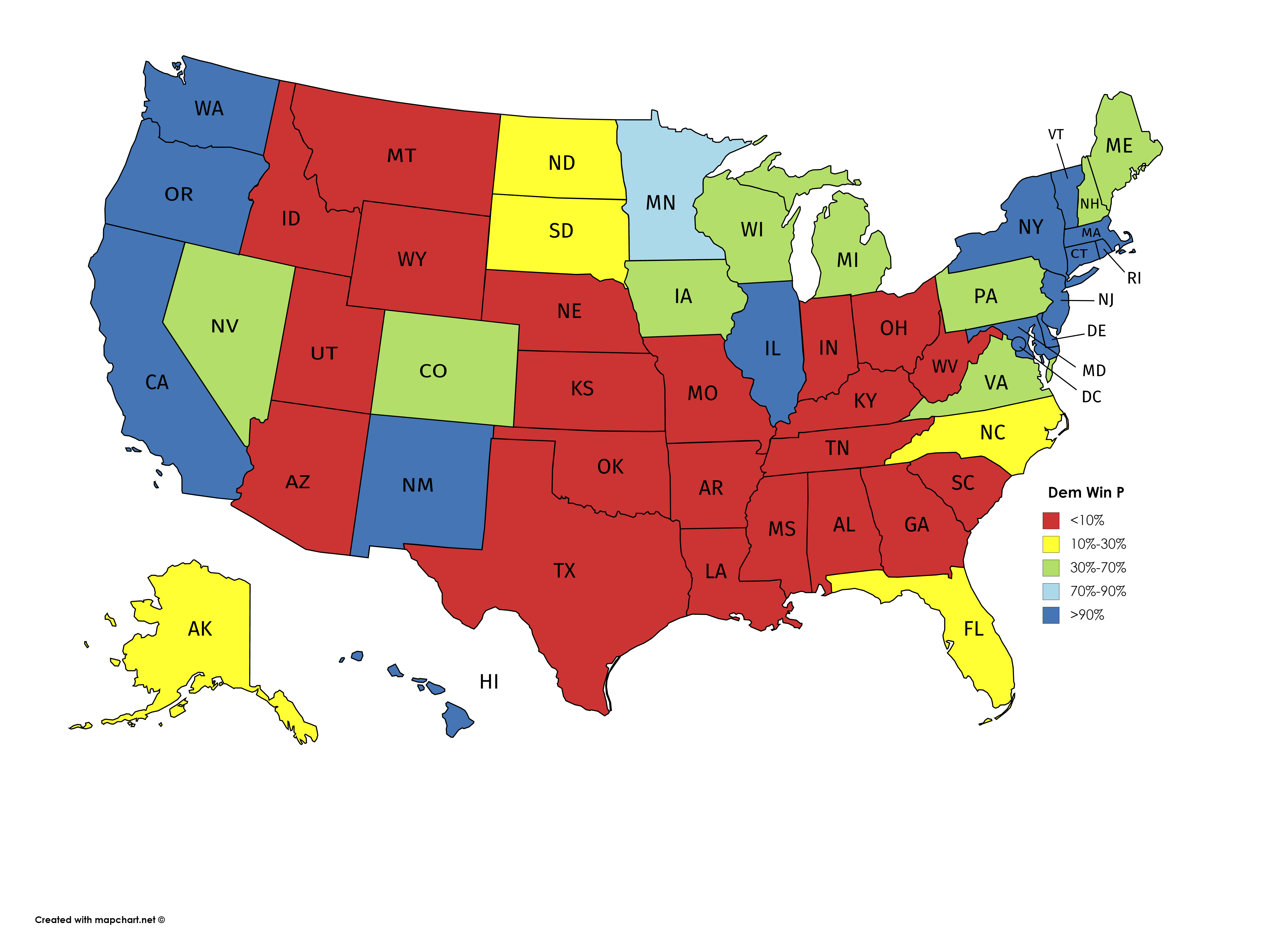}}
\caption{Electoral Maps for 2020 given nationwide PV}
\label{fig:maps}
\end{figure}

For each scenario, we repeat this simulation 10000 times. Probabilities of Democrats winning each state are in the file \texttt{stateProb.csv}. Histograms for EC votes are in Figure~\ref{fig:EC}. Red vertical lines show 269 votes: winning threshold. For each scenario, we compute the probability for Democrats winning EC (by getting more than 269 EC votes): Area to the right of the red line. Electoral maps for these 4 scenarios are given in Figure~\ref{fig:maps}. These maps are created using \texttt{mapchart.com}. We classify states by $p$ (the probability of Democrats winning) in 5 categories, similarly to the New York Times and other news agencies:

\begin{itemize}
\item $p > 90\%$: solid D, dark blue;
\item $70\% < p < 90\%$: lean D, light blue;
\item $30\% < p < 70\%$: swing states, green;
\item $10\% < p < 30\%$: lean R, yellow;
\item $p < 10\%$: solid R, red. 
\end{itemize}

\subsection{Evolution of the Electoral College bias} If we simulate the EC for even PV in 2012, 2016, 2020, then Democrats' win probability is 28\%, 31\%, 37\%, respectively. Thus the EC was, is, and will be biased in favor of Republicans. 

For previous elections: 2012 with actual 2012 PV, and 2016 with actual 2016 PV, we have Democrats' EC win probabilities 82\% and 61\%, respectively. Thus we see that the Democrats' lead in 2016 was less robust than in 2012. Compare this with 538 \texttt{fivethirtyeight.com}: 90.9\% and 71.4\% for final election predictions on the Election Day morning in 2012 and 2016. 

We have probabilities closer to 50\% because we rely only on past actual statewide votes in our backtesting, while 538 uses other pieces of information: polls, fundraising, endorsements.

Let us discuss whether the EC bias (deviation from 50.0\%) is due to statistical error. Assuming that the EC is, in fact, unbiased, the standard significance test for the fraction of simulations (out of $N = 10000$) in which Democrats win gives the interval $49.36\%--50.64\%$ for significance level $p = 0.05$. The results of 2012, 2016, 2020 fall outside this interval. This confirms that the EC is, in fact, biased towards Republicans. 

\subsection{State-by-state analysis} We summarize the table of state-by-state win probabilities in these 4 scenarios from \texttt{stateProb.csv}; see also  Electoral College maps in Figure~\ref{fig:maps}.

\begin{itemize}
\item The safest state for Democrats is California, with win probability greater than 99.9\% in all 4 scenarios. New York, New Jersey, Hawaii, Maryland, Washington, Connecticut are also very safe for Democrats, with probability greater than 99\%.

\item The safest state for Republicans is Oklahoma, with win probability greater than 99.9\%. Alabama, Idaho, Utah, Tennessee, Wyoming have win probability greater than 99\%. 

\item Even though the media discussed the idea of turning Texas blue, see for example \cite{Texas}, Democrats win Texas in 2020 for even or 2008 PV only with probabilities 0.8\% and 4.3\%, respectively. Thus the path to EC win for Democrats does not lie through Texas. Instead, they should follow a traditional path through Midwest and Northeast, trying to get back the states that Republicans won in the 2016 presidential election. We discuss this below, when we speak of state importance coefficients. 

\item The states with closest to 50\% win probability: Virginia for 2020 election with even PV; Nevada for 2020 election with 2016 PV; Ohio for 2020 election with 2008 PV; New Hampshire for 2020 election with 2004 PV.
\end{itemize}

\subsection{Electoral College maps for various scenarios} We analyze the maps from Figure~\ref{fig:maps}; the letters below correspond to subfigures in Figure~\ref{fig:maps}. 

\begin{enumerate}[label=(\Alph*)]
\item Except Nevada, all swing states in the Northeast and  Midwest: Virginia, Pennsylvania, Maine, New Hampshire, Iowa, Wisconsin. Although Ohio and Florida went D in 2008 and 2012, we rate Ohio solid R and Florida lean R. But Michigan is leaning D, and Wisconsin and Pennsylvania are rated swing. These three states were crucial for 2016 success of Donald Trump.  Interestingly, we rate Arizona and the Dakotas only lean R, despite them voting R in all presidential elections since 2000.

\item Since 2016 PV was in favor of Democrats, the EC is more Dem-favorable as compared to (A). Nebraska and Ohio change from solid R to lean R, Pennsylvania and New Hampshire change from swing to lean D, and North Carolina becomes swing.

\item Year 2008 was even more in favor of Democrats than 2016, thus the map in (C) shifts to Democrats compared to (B). Arizona, Florida, Indiana, North Dakota, Ohio switch to swing from lean R; Colorado and Pennsylvania switch from lean D to solid D; North Carolina, Virginia, Wisconsin switch from swing to lean D; Montana and South Carolina switches from solid R to lean R. 

\item Year 2004, on the contrary, was more favorable to Republicans. Thus the map becomes more R-leaning as compared to the map from (A):  Colorado and Michigan become swing states, Minnesota switches from solid D to lean D, and Arizona becomes solid R. 
\end{enumerate}

\newpage

\subsection{Pivotal states} Which states are important for victory? We discuss this from the viewpoint of Democrats; but since elections are a zero-sum game, the same analysis applies to Republicans. Assume Democrats can increase the $\alpha_s$ of this state to $\alpha_s + \mathrm{d}\alpha_s$. Then the probability of Democrats winning this state has changed
$$
\mbox{from}\quad\Phi\left(\sigma_s^{-1}(\alpha_s + \beta_sx(e) + \gamma_st)\right)\quad\mbox{to}\quad \Phi\left(\sigma_s^{-1}(\alpha_s + \mathrm{d}\alpha_s + \beta_sx(e) + \gamma_st)\right),
$$
or, in other words, by $\varphi(\alpha_s + \beta_sx(e) + \gamma_st, \sigma_s)\cdot\mathrm{d}\alpha_s$, with the normal density $\varphi(x, \sigma)$ defined by
$$
\varphi(x, \sigma) := \frac{\partial}{\partial x}\Phi(x/\sigma) = \sigma^{-1}\Phi'(x/\sigma) = \frac1{\sqrt{2\pi}\sigma}\exp\left(-x^2/2\sigma^2\right).
$$
Thus the expected rate of gain in Electoral College votes per change $\mathrm{d}\alpha_s$ is 
$$
\mathbb I_s := \varphi(\alpha_s + \beta_sx(e) + \gamma_st, \sigma_s)\cdot\mathrm{EC}_s,
$$
where $\mathrm{EC}_s$ is the number of EC votes in state $s$. This {\it importance coefficient} $\mathrm{I}_s$ is calculated for each state $s \in S$, for all 6 scenarios in the file \texttt{importance.csv}. In Table~\ref{table:importance}, we list the three most pivotal states together with their importance coefficients. 

\begin{table}
\begin{tabular}{|c|c|c|c|}
\hline
Scenario & Rank 1 & Rank 2 & Rank 3\\
\hline
2020 with even PV & PA (68) & NC (55) & WI (36)\\
2020 with 2016 PV & NC (73) & PA (55) & FL (40)\\
2020 with 2008 PV & OH (93) & PA (63) & IN (21)\\
2020 with 2004 PV & PA (64) & MI (44) & NC (32)\\
\hline
\end{tabular}
\bigskip
\caption{Most Important States and Importance Coefficients}
\label{table:importance}
\end{table}

\section{Conclusion}

We found the partisan lean of each state, its time dynamics, and its dependence of the national political environment, using Bayesian linear regression of statewide election PV vs election year and national PV. To test whether the EC is biased, we set the equal major party nationwide PV, and simulate the EC votes. It turns out that there is indeed a systematic and persistent bias of the EC in favor of Republicans. From Table~\ref{table:importance}, we see that the path to presidency lies through Pennsylvania, North Carolina, and the Midwest. 

Our version of PVI (partisan voting index) almost coincides with Cook PVI, and our back-casts for Presidential elections in 2012 and 2016 more or less match final predictions from FiveThirtyEight. Small states have significantly higher variance than large states. 

Again, we emphasize that this model is very simple. We do not pretend to describe voting behavior and its change over time in a comprehensive way. To this end, we need polls, ethnic and income composition data, etc. Some references are cited in the Introduction. 

Subsequent research might focus on more sophisticated models, which include ethnic and economic statewide data, or the power of incumbency. Other possible lines of research: (a) make sense of unusual elections which we disregarded in our analysis (see Section 2); (b) include state and local elections (county and city councils, governors, state legislatures, judges); (c) capture correlations between states with similar ethnic and economic patterns, such as Wisconsin and Michigan, or Nevada and Arizona; (d) study the size effect further (whether large state vote differently than small ones); (e) model Maine and Nebraska, which split their EC votes.

\end{document}